\documentclass[sigconf]{acmart}

\usepackage{acmart-taps}
\usepackage{dirtytalk}
\usepackage{xcolor}
\usepackage{listings}

\newcommand{\edit}[1]{#1}
\newcommand{\sys}{\textsc{AGDebugger}}

\definecolor{code}{HTML}{164E63}

\setcopyright{acmlicensed}
\copyrightyear{2025}
\acmYear{2025}
\setcopyright{cc} 
\setcctype{by}
\acmConference[CHI '25]{CHI Conference on Human Factors in Computing Systems}{April 26-May 1, 2025}{Yokohama, Japan}
\acmBooktitle{CHI Conference on Human Factors in Computing Systems (CHI '25), April 26-May 1, 2025, Yokohama, Japan}\acmDOI{10.1145/3706598.3713581}
\acmISBN{979-8-4007-1394-1/25/04}

\begin{document}
\title{Interactive Debugging and Steering of Multi-Agent AI Systems}

\author{Will Epperson}
\email{willepp@cmu.edu}
\orcid{0000-0002-2745-4315}
\affiliation{%
  \department{Human-Computer Interaction Institute}
  \institution{Carnegie Mellon University}
  \city{Pittsburgh}
  \state{PA}
  \country{USA}
}

\author{Gagan Bansal}
\email{gaganbansal@microsoft.com}
\orcid{0000-0002-7741-3861}
\affiliation{%
  \institution{Microsoft Research}
  \city{Redmond}
  \state{WA}
  \country{USA}
}

\author{Victor Dibia}
\email{victordibia@microsoft.com}
\orcid{0000-0002-1839-5632}
\affiliation{%
  \institution{Microsoft Research}
  \city{Redmond}
  \state{WA}
  \country{USA}
}

\author{Adam Fourney}
\email{adam.fourney@microsoft.com}
\orcid{0000-0002-4986-7794}
\affiliation{%
  \institution{Microsoft Research}
  \city{Redmond}
  \state{WA}
  \country{USA}
}

\author{Jack Gerrits}
\email{jagerrit@microsoft.com}
\orcid{0009-0003-1966-2434}
\affiliation{%
  \institution{Microsoft Research}
  \city{Redmond}
  \state{WA}
  \country{USA}
}

\author{Erkang Zhu}
\email{erkang.zhu@microsoft.com}
\orcid{0009-0000-3326-1790}
\affiliation{%
  \institution{Microsoft Research}
  \city{Redmond}
  \state{WA}
  \country{USA}
}

\author{Saleema Amershi}
\email{samershi@microsoft.com}
\orcid{0000-0002-3294-7288}
\affiliation{%
  \institution{Microsoft Research}
  \city{Redmond}
  \state{WA}
  \country{USA}
}

\renewcommand{\shortauthors}{Epperson et al.}

\begin{abstract}
Fully autonomous teams of LLM-powered AI agents are emerging that collaborate to perform complex tasks for users.
{\em What challenges do developers face when trying to build and debug these AI agent teams?}
In formative interviews with five AI agent developers, we identify core challenges: difficulty reviewing long agent conversations to localize errors, lack of support in current tools for \textit{interactive} debugging, and the need for tool support to iterate on agent configuration.
Based on these needs, we developed an interactive multi-agent debugging tool, \sys{}, with a  UI for browsing and sending messages, the ability to edit and reset prior agent messages, and an overview visualization for navigating complex message histories.
In a two-part user study with 14 participants, we identify common user strategies for steering agents and highlight the importance of interactive message resets for debugging. 
Our studies deepen understanding of interfaces for debugging increasingly important agentic workflows.
\end{abstract}



\begin{teaserfigure}
  \centering
  \includegraphics[width=.8\textwidth]{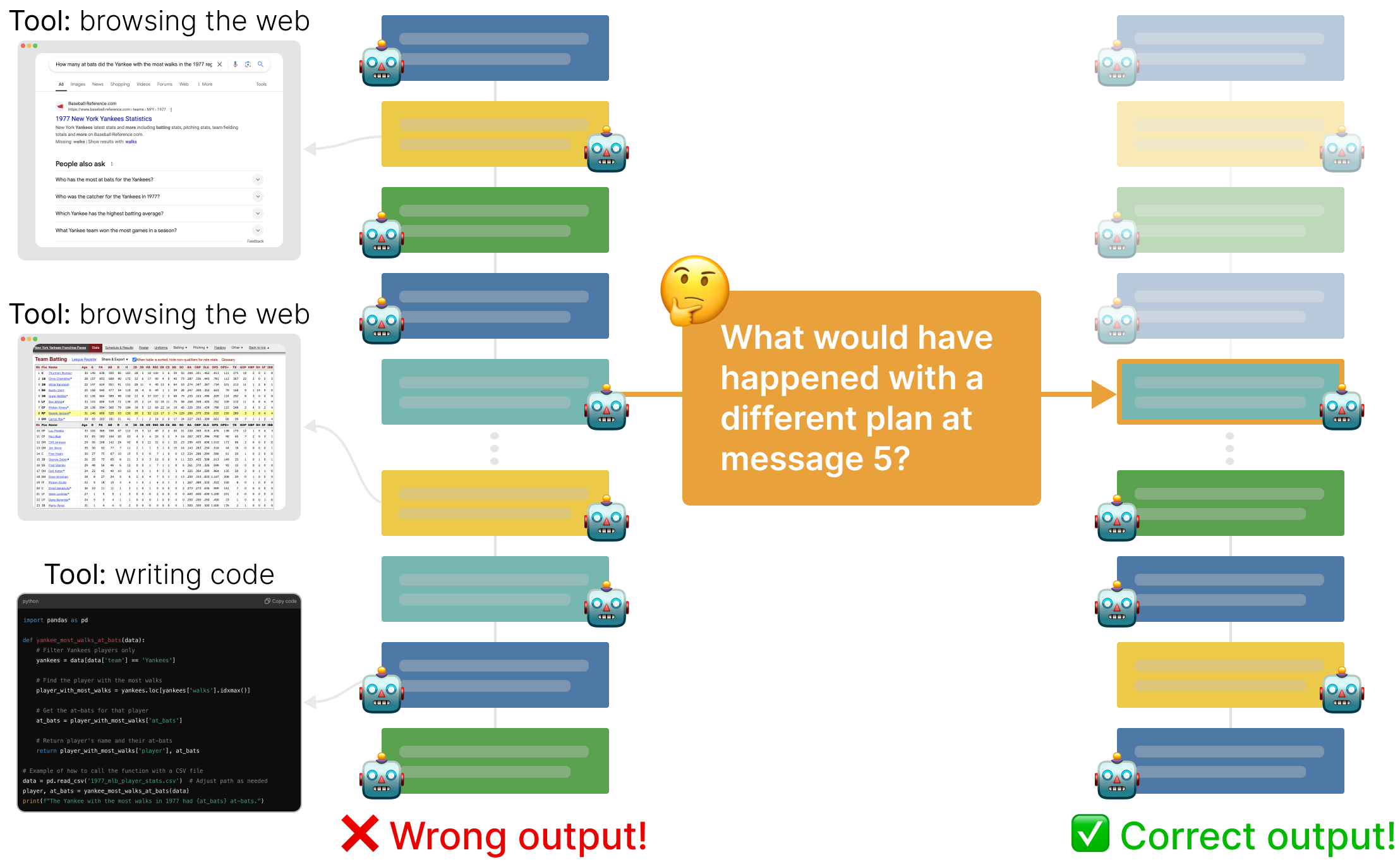}
  \caption{
  Debugging multi-agent AI systems involves reasoning over long multi-turn conversations where specialized agents use tools like web browsing and writing code with LLMs. 
  \sys{} allows users to interactively debug and steer multi-agent teams by resetting the agents to earlier points in the workflow then editing messages to interactively test hypotheses about their behavior.
  }
  \Description{This figure shows an agent conversation with represented as boxes on the lefthand side. Several of these boxes are connected to screenshots of the tool used in that box such as a web browsing page or code writing. One of the messages is annotated with an annotation that reads What would have happened with a different plan at message 5?. This leads to a second conversation that is a fork of the first where the messages above the fork are greyed out and the new messages are different. At the bottom, an annotation shows that the original agent conversation led to the wrong output with a red x, but the new one leads to the correct output with a green check.}
  \label{fig: agdebugger teaser}
\end{teaserfigure}

\maketitle

\section{Introduction}

Multi-agent AI systems are emerging as a powerful paradigm for addressing tasks beyond the scope of single large language models (LLMs)~\cite{wu2024autogen, Li2023CAMELCA, wu2024oscopilot, crewai2024}. 
By combining LLMs with multi-turn state tracking, external tool use, and collaborative interactions, multi-agent systems can perform complex, real-world tasks such as accessing up-to-date information or executing actions in dynamic environments~\cite{he-etal-2024-webvoyager}. 
This capability has allowed multi-agent AI systems to excel in challenging benchmarks that require reasoning about complex tasks and using tools to interact with the world like web browsing~\cite{he-etal-2024-webvoyager}, reasoning over complex files~\cite{gaia2024}, and writing and executing code~\cite{jimenez2024swebench, wang2024opendevin}.

Despite these advances, existing AI development tools fall short when it comes to understanding and debugging the complex, multi-turn behaviors of agent teams. 
Traditional AI debugging practices, which focus on model training or correcting datasets, are inadequate in this new paradigm. 
Much of the work in debugging LLM systems centers around crafting effective text \textit{prompts} that instruct the LLM how to accomplish a task through in-context learning~\cite{LLM_few_shot_2020, jonnyCantPromptZam2023}.
Prior research has developed tools for crafting effective prompts for tasks involving individual LLM invocations~\cite{Jiang2022PromptMakerPP, tenney2024sequencesalience, petridisConstiMaker2024, openAiPlayground} or chains of LLM calls~\cite{wuPromptChainer2022, arawjo2023chainforge}, but has not addressed the challenges associated with debugging teams of fully-autonomous AI agents.

\edit{
Debugging multi-agent teams introduces new debugging challenges since agent teams require first crafting individual prompts and tools for each agent, then understanding how the team works together to accomplish a task by making numerous LLM calls over a multi-turn conversation. 
Agent conversations are complex and dynamic, where agents formulate a plan on the fly for a task and then execute the plan by using tools to interact with the world as needed~\cite{wu2024autogen, crewai2024}.
Debugging multi-agent systems requires simultaneously understanding both the individual behaviors of agents and the emergent interactions between them. 
This multifaceted debugging challenge demands new tools that can integrate insight across the entire agent team to avoid \say{cascading errors} that fix one component while breaking another~\cite{wuPromptChainer2022, mltechdebtSculley2014, nushi2017troubleshooting}.} 

While recent systems like AutoGen Studio~\cite{dibia2024AutogenStudio} and OpenDevin~\cite{wang2024opendevin} enable developers to interact with multi-agent AI teams, they primarily focus on task execution and agent construction. 
These platforms lack robust debugging features, particularly for multi-turn interactions. 
As agent conversations grow longer, existing tools do not provide the ability to pause, rewind, or edit agent behaviors in real time.
Additionally, they lack comprehensive visualizations to help developers track and understand the evolving dynamics between agents. 
These gaps in literature makes debugging multi-agent systems particularly challenging, as developers need to iteratively diagnose and adjust interactions across the team.

To address the limitations of current tools and better support the debugging of multi-agent AI systems, we explore two key research questions:
(1) How can we design systems that enable developers to effectively debug multi-agent AI teams?
(2) How do developers use such a system to debug and improve agent workflows in practice?

To investigate these research questions, we first conducted formative interviews with five expert developers who have extensive experience building multi-agent AI systems. 
These interviews revealed several key challenges in the debugging process, such as difficulties in understanding long, multi-turn agent conversations, the lack of interactive debugging support in existing tools, and the need for better tooling to iterate on agent configurations.

Using these findings, we designed an interactive debugging system, \sys{} to address these challenges. 
\sys{} has three primary features to facilitate the debugging process.
The first builds off prior interfaces for agent configuration~\cite{dibia2024AutogenStudio, wang2024opendevin, langgraphStudio2024} and allows users to interactively send messages to the agents and inspect the history of messages sent with fine-grained control on the execution of messages.
An extension to chatting with agents is the ability to interactively control a conversation by resetting to previous points and editing previously sent agent messages.
\sys{} enables such an interaction by check-pointing agent state (including state that might be impacted by agent actions and tool use) before each message is sent to enable resetting to earlier points in a conversation and editing agent messages.
Finally, as agent conversations grow longer and users edit the conversation history they can become difficult to track. 
\sys{} also includes an interactive overview visualization for summarizing the agent conversations and edits.

We conducted a two-part user study with 14 participants to evaluate the effectiveness of \sys{} for debugging multi-agent workflows. 
In the first part, participants diagnosed errors in agent workflows by using \sys{} to inspect and understand where the agents failed. 
In the second part, they used \sys{} to edit agent messages and steer the agents toward successful outcomes. 
Our findings reveal that participants frequently made three types of modifications: specifying more detailed instructions, simplifying agent tasks, and altering the agents' plans. 
These patterns reflect common failure modes in multi-agent systems, and \sys{} effectively supported participants in iteratively steering the agents towards correct behavior. 
Finally, we conclude by presenting remaining open challenges in agent debugging found from our study. 
Such challenges include decoupling steering from the agent implementation and determining if edits had an effect.
\sys{} is available as an open source tool at \url{https://github.com/microsoft/agdebugger}.
In summary, this paper makes the following contributions:
\begin{enumerate}
    \item Formative interviews with agent developers that reveal common challenges developers encounters while developing multi-agent AI systems including understanding long, multi-turn agent conversations, the lack of interactive debugging support, and the need for better tooling to iterate on agent configurations.
    \item A prototype agent debugging system, \sys{}, with three key features for facilitating agent debugging: interactively sending and stepping through agent messages, the ability to interactively edit and steer agent teams, and an overview visualization to summarize agent conversations and edits.
    \item Results from a user study where participants use \sys{} to diagnose and then experiment with fixes for errors in agent workflows. We identify common patterns of steering across users around adding instructions to agent messages, simplifying messages, and modifying the agents' plan.
\end{enumerate}

\section{Related Work}

\subsection{Multi-Agent AI Systems}

In recent years, general purpose Large Language Models (LLMs) leveraging in-context learning (e.g., few-shot prompting), have been applied to various domains that previously required training specialized models~\cite{LLM_few_shot_2020}.   
When combined with the ability to use tools and store state, these models can function as agents that interact with the world to accomplish more complex tasks.
AI agents have been developed for tasks ranging from browsing the web~\cite{iclrZhouWebArena2024, he-etal-2024-webvoyager} to advanced coding tasks that involve looking up documentation and combining code authoring with execution~\cite{jimenez2024swebench, wang2024opendevin}.

For more intricate problems, performance improvements can often be achieved by \textit{multi-agent systems}, where multiple specialized agents work together to plan and perform tasks~\cite{wu2024autogen, Li2023CAMELCA, wu2024oscopilot, guo2024large}. 
\edit{These agents use LLMs for decision making and planning by breaking down tasks into smaller sub-components, leverage tools to interact with the world, and memory to keep track of past actions and provide context for subsequent actions~\cite{guo2024large}.
Multi-agent systems typically combine multiple \textit{specialized} agents, each with different roles, prompts, tools, and memory that collaborate together to solve tasks by planning and delegating to the appropriate agent.
Such systems demonstrate strong performance on complex tasks and make it easier for developers to design and develop reusable agents for different tasks, similar to object oriented programming~\cite{cheng2024exploring, guo2024large}.
Creating multi-agent systems involves configuring each individual agent's capabilities and tools as well as how agents communicate with one another.
Frameworks such as AutoGen~\cite{wu2024autogen}, CAMEL~\cite{Li2023CAMELCA}, OS-Copilot~\cite{wu2024oscopilot}, Crew AI~\cite{crewai2024}, and LangGraph~\cite{langgraphStudio2024} facilitate the development of such systems, enabling the creation of LLM-powered agents with distinct tools and roles.}

Despite these advances, LLMs exhibit various failure modes, such as losing track of critical information over long contexts~\cite{liu-etal-2024-lost} and hallucinating facts that lack contextual grounding~\cite{maynez-etal-2020-faithfulness}. 
These underlying model issues remain in multi-agent settings and can be exacerbated when agents interact through chained model calls, further complicating error tracking and resolution~\cite{2022-ai-chains, wuPromptChainer2022}.

\subsection{LLM and Agent Debugging}
Debugging a single LLM component or team of agents builds on prior research on debugging traditional machine learning and AI models.
Researchers have previously recognized the difficulty of diagnosing model failure modes~\cite{amershiModelTracker2015, sweForMLAmershi2019}. 
In classic ML workflows, interactive debugging tools have been developed to assist developers at various stages of model development, from tracking model changes~\cite{amershiModelTracker2015} to identifying poorly performing data subsets during evaluation~\cite{cabreraZeno2023}. 
One key insight from these tools is the value of providing developers with direct access to the raw data driving model decisions, enabling them to better understand and intervene when failures occur~\cite{patel2010gestalt, cabreraZeno2023}.

Prior research also discusses the value of \textit{counterfactual} explanations for understanding ML models~\cite{counterfactualWacher2017}.
Such methods allow users to explore \say{what if} scenarios by modifying model inputs and observing how predictions shift~\cite{wexlerWhatIfTool2020, limWhy2009, kuleszaExplanatory2015}.
Interactive systems support what if analysis of NLP models, aiding users in generating variations of input sentences to test how model outputs change~\cite{wu2021polyjuice, ribeiro-lundberg-2022-adaptive, cheng2024InteractiveCounterfactual}.

For LLMs, debugging primarily involves crafting effective prompts to specify a task, rather than selecting model training data or architectures. 
Prompt engineering has become central to eliciting desired LLM behaviors, but studies show that users often struggle to express their design goals in prompts, and face challenges debugging incorrect model outputs~\cite{Jiang2022PromptMakerPP, jonnyCantPromptZam2023}. 
Interactive tools allow users to experiment with different system prompts and models~\cite{openAiPlayground, strobelt2022promptIDE}.
\edit{Tools like PromptMaker~\cite{Jiang2022PromptMakerPP} and Sequences Salience~\cite{tenney2024sequencesalience} aim to assist users in refining their prompts, while other systems allow for iterative feedback-driven adjustments~\cite{petridisConstiMaker2024}. 
Other systems help users compare the outputs from multiple prompts and use a \say{LLM as a judge} to grade which outputs are better~\cite{kahng2025llmcomparator, kimEvalLM2024, shankarValidate2024} or unit test LLM components with code or model-based assertions~\cite{promptfoo2024, autoblocksAI2024}.
Recent research like ChainForge~\cite{arawjo2023chainforge} or PromptChainer~\cite{wuPromptChainer2022} have contributed tools for creating and debugging pipelines of LLM calls, however do not support LLM based agents.} 

\edit{Tools focused on debugging a single LLM component or pre-defined pipeline of LLM calls do not fully address the challenges associated with debugging multi-agent systems. 
Muti-agent debugging requires understanding both individual agent behavior, tool use, and memory as well as agent interactions over long multi-turn conversations including appropriate delegation or task termination.
Improvements in one component can introduce errors in others, underscoring the need for testing the entire agent team simultaneously on a task---a problem previously identified for pipelines of traditional AI systems or chains of LLM calls~\cite{mltechdebtSculley2014, amershiModelTracker2015, nushi2017troubleshooting, wuPromptChainer2022}.} 

\edit{Recent systems like AutoGen Studio~\cite{dibia2024AutogenStudio}, OpenDevin~\cite{wang2024opendevin}, or Crew AI~\cite{crewai2024} provide interfaces for creating and interacting with multi-agent AI teams for a task, allowing developers to communicate with agents via a chat-style UI. 
While these systems support agent construction, they focus less on the interactions required for interactive debugging. 
There remains limited HCI research on the design of tools for debugging multi-agent systems and how developers use such tools.}

\edit{Prior research has demonstrated the power of pause and reset mechanisms for debugging, such as the crash-and-rerun programming model of TurKit that allows programmers to re-run programs that make expensive function calls to crowdworkers~\cite{littleTurkit2010}.
Interactive tools for LLM debugging like the OpenAI Playground provide the ability to edit earlier LLM messages, but only support interaction with a single LLM in a chat rather than a multi-agent team~\cite{openAiPlayground}.
LangGraph offers a UI for creating multi-agent systems based on a graph communication model and offers support for breakpoints to inspect agent state when called in the graph~\cite{langgraphStudio2024}.
However their implementation is not public, requires developers to pre-specify breakpoints, and only works for graph-based agents.
The pause and reset interactions presented in \sys{} are applicable to any multi-agent system that interacts by passing messages.}

\section{Background: Agent Framework and Tasks}
\label{sec: background}

This section describes three key background concepts: the implementation of the agent framework we used, reviews the GAIA benchmark dataset for evaluating AI agents~\cite{gaia2024}, and details an example of a specific multi-agent team (implemented using this framework) that achieves high performance on GAIA.

\subsection{Agent Implementation Framework}

Our debugging system is built on the open-source AutoGen framework~\cite{wu2024autogen}. 
In this framework, agents are implemented as Python classes that communicate by sending {\em messages} through a shared {\em runtime}. 
These messages exchanged are also typed Python objects containing data.
Agents implement {\em message handlers} that respond to particular types of messages.
When an agent receives a message, the framework triggers the appropriate handler, which might make LLM calls, use tools, and send a new message in a response. 
Furthermore, while processing a new message, agents often update their state such as navigating to a new page in a web browser the agent controls.
All messages are sent through the central runtime which manages a {\em message queue}. 
When messages are processed, they are moved off the queue and sent to the appropriate agents. 
This design allows for flexible patterns of communication between agents, where agents can communicate directly with each other or send messages to all other agents.

In addition to messages, agents can have internal \say{thoughts} that are simply log messages. These thoughts are not sent to other agents but can be helpful for debugging. 
Each agent can keep track of its message history to provide context for future model calls.

\subsection{GAIA Benchmark Tasks}

An agent framework by itself is only half the story: it needs to be applied to tasks or work to be of value. In this paper, we apply the framework to problems from the GAIA agent benchmark, allowing us to measure the proficiency of our agents and teams.
The GAIA benchmark is a collection of challenging AI assistant tasks that require diverse skills such as coding, using the internet, and parsing files~\cite{gaia2024}. 
It serves as a standardized way to evaluate the capabilities of AI agents across a range of complex tasks.
GAIA tasks are divided into three levels of difficulty, with level 1 the easiest and level 3 the hardest. 
As of writing, these tasks remain very challenging for AI assistants, with the top-performing team on the GAIA leaderboard scoring an average of around 35\% on the test set~\cite{gaia_leaderboard2024}.

In our studies and examples, we focus on two specific tasks from the validation set of GAIA Level-1, shown in Table~\ref{tab: study tasks}. 
These tasks are complex as they involve searching for and synthesizing information from multiple websites to generate the final answer.
While our agent team performs near the state-of-the-art on the GAIA leaderboard, it consistently fails to complete these specific tasks correctly, making them ideal candidates for debugging.
On both tasks, our agent team outputs an answer, but the answers are incorrect.
We investigate these two tasks specifically in our user study since both tasks are similar (e.g. web-focused and same difficulty level) with prompts that are easy for participants to understand without extra background.

\begin{table}[ht]
    \centering
    \begin{tabular}{cp{5cm}p{1.5cm}}
    ID & Benchmark question & Answer \\
    \hline
    T1 & How many at bats did the Yankee with the most walks in the 1977 regular season have that same season?
    & 519\\
    \hline
    T2 & Of the cities within the United States where U.S. presidents were born, which two are the farthest apart from the westernmost to the easternmost going east, giving the city names only? Give them to me in alphabetical order, in a comma-separated list
    & Braintree, Honolulu\\
    \end{tabular}
    \caption{Example agent tasks for debugging from the GAIA Level-1 validation set~\cite{gaia2024}.}
    \label{tab: study tasks}
\end{table}

\subsection{Agent Team for GAIA Tasks}

To address the GAIA benchmark tasks, we use the Magentic-One generalist AI agent team~\cite{fourney2024magentic-one}. 
This team, implemented using the framework described above, consists of five agents that collaborate to solve tasks:

\begin{enumerate}
    \item An \textit{Orchestrator} who plans and controls the conversation
    \item A \textit{Coder} who authors Python code to solve subproblems
    \item An \textit{Executor} who executes code locally and returns results
    \item A \textit{File Surfer} who can parse and interact with local files of various formats (e.g., PDF, PowerPoint, etc.)
    \item A \textit{Web Surfer} who can access and interact with web pages in a browser, and can perform search queries, in a manner comparable to prior web agents~\cite{he-etal-2024-webvoyager}
\end{enumerate}

Each of these agents maintains their own state, has access to different tools to fulfill user requests, and can make their own calls to LLMs. 
To complete a task, the agents are given an input prompt, then collaboratively develop and execute a plan to solve the task and produce a final result. 
These agent conversations can become quite lengthy. 
For example, in the runs we analyze, it took the agents 71 messages to produce an answer for task T1, and 90 messages for task T2. 
The raw log files with the messages contain 6,368 words for T1 and 7,230 words for T2.
This complexity in agent interactions highlights the need for effective debugging tools, which is the focus of our research.

\section{Formative Interviews on Agent Debugging}
\label{sec: formative interviews}

To better understand current developer pain-points around developing multi-agent AI systems, we interviewed five developers at Microsoft with experience building multi-agent applications.
We recruited these participants within a large technology corporation.
All five had prior experience using the AutoGen multi-agent framework~\cite{wu2024autogen}: two were core contributors to the open source project and the other three had experience developing multi-agent prototypes with the framework.
\edit{Our participants were three research scientists, one software engineer, and one engineering manager.}

We conduced semi-structured interviews in a one hour session to ask each participant about their development experience building multi-agent systems.
In our interviews, we asked each participant questions about their prior experience developing agents, challenges they ran into, and desired features from agent debugging tools.
Our exact interview questions are included in Appendix~\ref{appendix: formative interview details}.
We took detailed notes during each interview and then did thematic analysis of interview notes to synthesize common themes. 
Our interviewees described three primary pain points in developing multi-agent apps, detailed below.

\subsection{Understanding Long Agent Conversations is Cumbersome}
The first pain-point that our participants discussed is the difficulty of understanding long agent conversations.
To understand the results of a workflow, participants currently write all the messages exchanged between agents to the system console. 
The console is then saved as a single output text file and then reviewed post-hoc.

For a single task, on the order of 50-100+ text-heavy messages might be exchanged between agents. Each message has metadata information like the sender of the message, its recipient, and any text generated from LLM calls or tool invocations. In some cases, individual messages can themselves be difficult to interpret (e.g., the output of Python scripts written by agents). However, interpretation challenges compound as conversations grow and more agents are involved.
Participants must read lengthy histories to understand both how the agents act and where things might be going wrong. 

\subsection{Lack of Support for Interactive Debugging}
The next pain point described by participants is the current lack of support for an interactive debugging experience.
Participants desired the ability to have fine-grained control over the agents as they progress through a task.
This includes interrupting the agents if they seem to be stuck or going down the wrong path, resetting agents to earlier points in the conversation, and editing messages to steer agents towards the desired goal.
To this end, both P2 and P4 mentioned their desire to use \say{breakpoints} for agent debugging. Once reached, breakpoints would interrupt agent execution, and allow developers to understand and manipulate the state of each agent before continuing.
For example, PDB (python debugger) provides a comparable experience for debugging traditional python scripts~\cite{python_pdb_2024}.
Likewise, P5 drew parallels with the difficulties of debugging distributed systems, and expressed a desire for tools that capture, replay, and step through agent messages one at a time, step by step. This style of interaction would be helpful because issues often occur midway through a workflow. For instance, participants described how the AI agents often suggest reasonable plans, execute the first few steps of the plans correctly, then get stuck on a particular step, throwing off the rest of the workflow. In such cases, participants expressed a desire to reset the workflow to the last point where progress was being made, and then retry from this location---perhaps with a newly-corrected plan.

\subsection{Iterating on Agent Configuration}
Finally, four participants noted that iterating on agent configurations is currently a slow and arduous process. While debugging, developers are continuously tweaking their agent configurations by changing the system prompts, adding or removing agents from the team, or altering the selection of available tools. At present, developers must restart the workflows from the beginning to test the effectiveness of any given change. In cases where errors arise later in the conversation, developers must then wait considerable time to observe any impacts. Moreover, due to the stochastic nature of LLMs, the same errors might not always occur, requiring multiple run-throughs to gain confidence in a remediation. All of this slows down the debugging process considerably. To this end, participants expressed a desire to \say{freeze} the conversations at critical points and then iterate on potential fixes while the problematic context is isolated and in memory.

\subsection{Design Goals}
\label{sec: design goals}

Based on our formative interviews and the reviewed studies on AI debugging, we synthesized the following design goals for an interactive debugging tool for AI agents.
Users with such a tool should be able to:

\begin{enumerate}
    \item [G1.] \textbf{Understand messages exchanged between agents.} An agent debugging tool needs to expose the messages sent between agents so that users can understand the details of the conversation and how the agents are progressing through tasks. This is important for identifying \textit{where} errors are happening in the workflow.
    \item [G2.] \textbf{Interrupt the conversation and send new messages.} Users should be able to  pause/interrupt the workflow at any point, and send new messages to the agents. 
    \item [G3.] \textbf{Reset back to a previous point in the workflow} Once a failure point is identified, users need the ability to reset to an earlier point in the workflow in order to experiment with steering agents to alternate paths.
    \item [G4.] \textbf{Change agent configurations.} An agent debugging tool should let users change agent configurations, such as the prompts or models used, in order to experiment with fixes.
\end{enumerate}

\begin{figure}[ht]
    \centering
    \includegraphics[width=\linewidth]{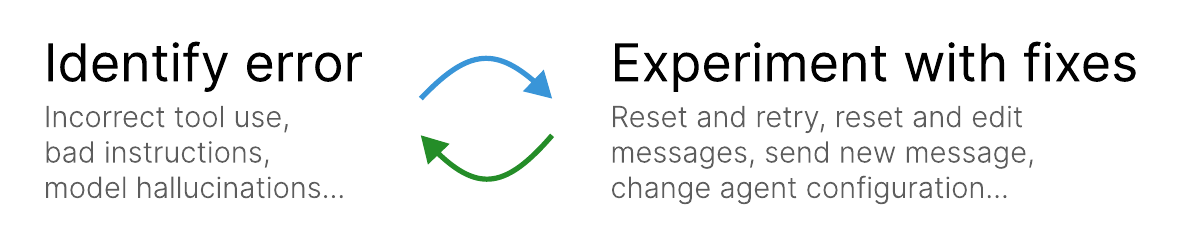}
    \caption{The agent debugging loop where developers iteratively identify errors and experiment with fixes that shape their understanding of the issue.}
    \Description{This shows a loop with identifying error on one side and experiment with fixes. Under identify error subtext reads incorrect tool use, bad instructions, model hallucinations… Under experiment with fixes subtext reads Reset and retry, reset and edit messages, send new message, change agent configuration...}
    \label{fig:debug loop}
\end{figure}

These design goals are aimed at supporting the debugging loop presented in Figure~\ref{fig:debug loop}, which is inspired by traditional software and AI debugging loops. 
An AI agent debugging tool should be able to help users both identify errors and experiment with fixes.
\textbf{G1} deals with error identification, \textbf{G4} deals with experimenting with fixes, and \textbf{G2} and \textbf{G3} support both identification and experimentation. 
We view this debugging process as a self-reinforcing loop where users identify errors, then experiment with fixes which allows them to refine their understanding of the error.

\section{\sys{}: Interactive Agent Debugging}
\label{sec: system description}

Based on our design goals, we developed \sys{}, an interactive debugging tool for AI agents.
\sys{} has three core features that enable interactive agent debugging:
(1) it presents the messages exchanged between agents in an interactive message viewer that also lets users send new messages,
(2) it lets users reset the agents workflow to earlier points in the conversation to edit,
and (3) it has an overview visualization for helping users navigate long conversation histories.

\subsection{Message Sending and History}
\label{sec: feat -- message viewer}

The first feature of \sys{} is the ability to send new messages to agents and view the messages exchanged between agents (design goals \textbf{G1} and \textbf{G2}).
Users can send messages to start the agents working on a new task or pause agent execution to send new messages during the middle of a run.
\edit{This type of interaction draws design inspiration from standard AI chat apps and prior agent creation and chat interfaces~\cite{dibia2024AutogenStudio, crewai2024, langgraphStudio2024}}.
Our message sending feature extends these designs by providing fine grained conversation control with the ability to pause execution and send new messages to particular agents in the middle of a conversation.

\begin{figure*}[t!]
    \centering
    \includegraphics[width=.83\linewidth]{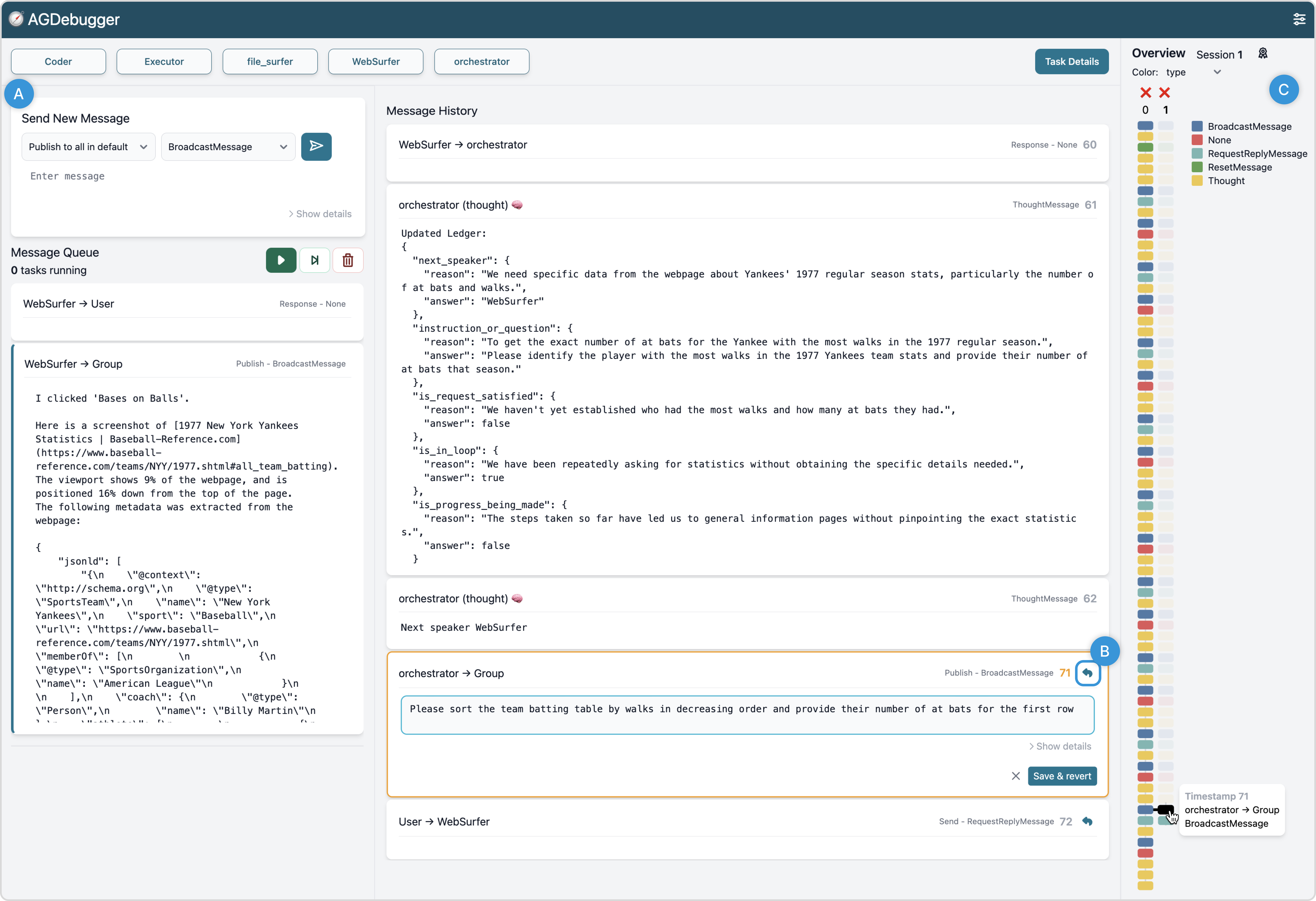}
    \caption{\sys{} helps users interactively debug and steer their agent teams. (A) Users can interactively send new messages, control the flow of messages, and see the history of agent messages 
    (Section~\ref{sec: feat -- message viewer}).
    (B) Users can revert to earlier points in the workflow by resetting and editing messages (Section~\ref{sec: feat -- reset messages}). 
    (C) The overview visualization helps users make sense of long conversations and the history of edits in an interactive visualization (Section~\ref{sec: feat -- overview visualization}). 
    }
    \Description{This shows a screenshot of the AgDebugger UI with three columns. On the left and annotated with (A), is a message sending panel with a text entry box and message sending controls. Under it I the message queue with 2 messages in the queue. In the middle is the message history view which shows 6 messages exchanged between agents and their contents. One of the messages is annotated with a B and shows the interactive resetting and editing features. On the right is the overview visualization showing a rectangle for all the messages sent in the conversation with a fork point after a reset.}
    \label{fig:system}
\end{figure*}

Figure~\ref{fig:system} shows the \sys{} interface,  with message sending controls appearing in panel A.
This feature lets users send a new message to all other agents (broadcast) or to any one agent in particular. 
When a message is sent it is appended to the message queue, where all messages are processed in the order in which they arrived. Once processed, the message is recorded in the message history, along with an execution timestamp.
The message queue can be run automatically with the play button or can be stepped through one message at a time by the user.
This step-by-step debugging is similar to line-by-line execution provided by python debuggers like PDB~\cite{python_pdb_2024}.

In the example in Figure~\ref{fig:system} we can see the history of the current conversation in the message history panel with the most recent messages at the bottom.
Here, the Orchestrator has asked the Web Surfer to sort a table in the current web page, and the Web Surfer has initiated an internal monologue (thought) to determine which low-level action(s) will accomplish the Orchestrator's instruction.
The next message in the queue is the result of the Web Surfer's action where it reports on clicking on a particular part of the page.

\subsection{Message Resetting and Edits}
\label{sec: feat -- reset messages}

While the ability to send messages to agents and view the conversation history is included in prior agent configuration tools, \sys{} provides a novel interaction with the ability to reset agents to previous points in the conversation and make edits to messages (design goal \textbf{G3}). 
\edit{
Since agents are stateful, resets and edits require more than a simple transformation of the message history (e.g. truncating the transcript) that might work for non-agentic LLM applications. 
\sys{} offers robust checkpointing support to reset the states of the agents themselves (e.g., having the Web Surfer return to the web page it was visiting at that time).}

Users can reset to an earlier point in a workflow in two ways.
They can directly edit a historical message inline then save the edit to reset the conversation back to this timestamp.
Or users can click the reset button on a message if they wish to restore the conversation back to that point without any changes to the message (e.g., to simply retry the flow).
This resetting interaction provides an affordance for users to ask two core agent debugging questions:
\begin{enumerate}
    \item What happens if I retry the workflow from this point?
    \item What \textit{would have happened} if this alternative message had been produced?
\end{enumerate}

\begin{figure}[t!]
    \centering
    \includegraphics[width=.9\linewidth]{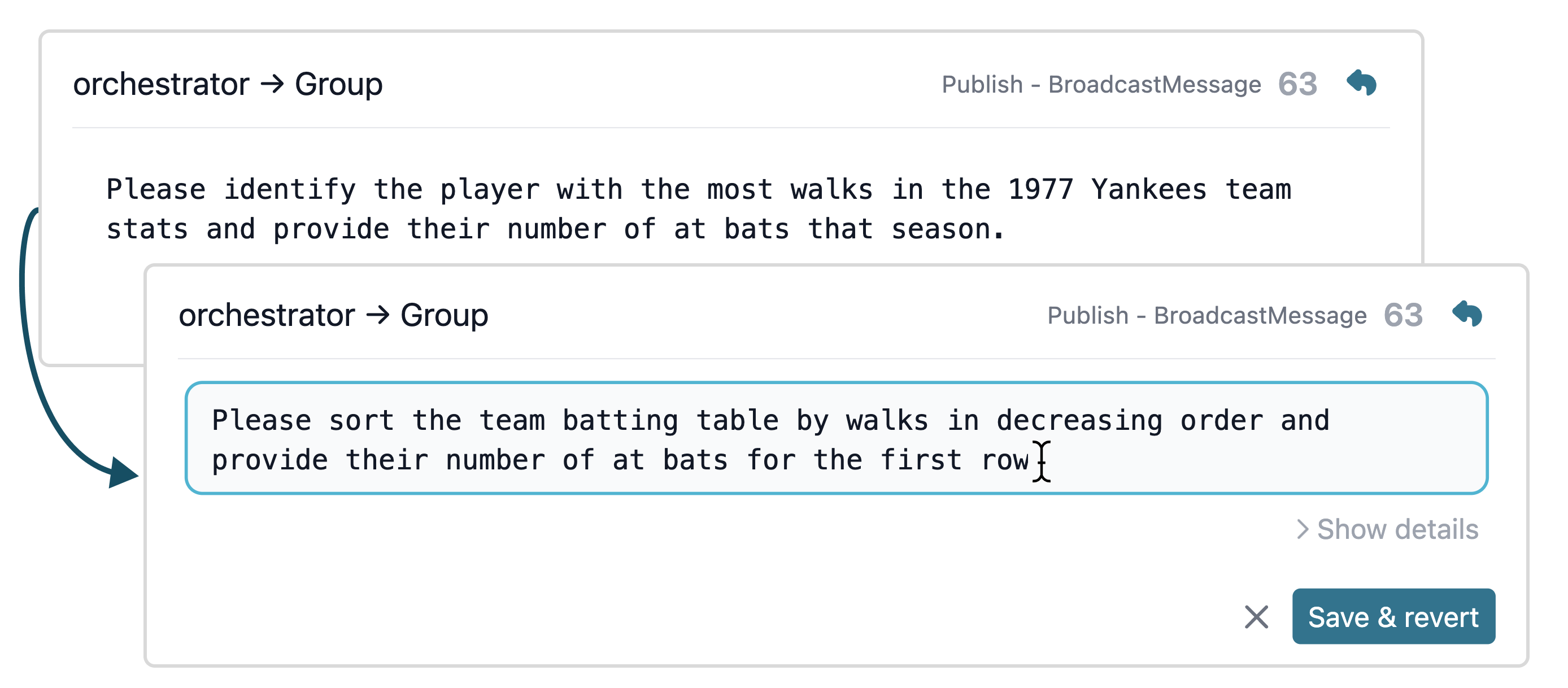}
    \caption{
    Users debug agent workflows by directly editing prior agent messages then restarting the workflow from that point, such as adding more specific instructions to a message to steer the agents towards the correct outcome. }
    \Description{A user edits a message to change the text from "Please identify the player with the most walks in the 1977 Yankees team stats and provide their number of at bats that season" to "Please sort the team batting table by talks in decreasing order and provide their number of at bats for the first row"}
    \label{fig:reset}
\end{figure}

Figure~\ref{fig:reset} shows an example where one agent is requesting information from another agent, and the user is leveraging the edit and reset capabilities to refine the request by providing more specific instructions.
This gives the user the ability to see \textit{what would have happened} if the Orchestrator had produced a different plan.
If the workflow now succeeds, they know to focus their efforts on making the plans more precise rather than perhaps tweaking how the web agent executes the plans.

\subsubsection{Technical details: checkpoints and sessions}
\begin{figure}[t!]
    \centering
    \includegraphics[width=.9\linewidth]{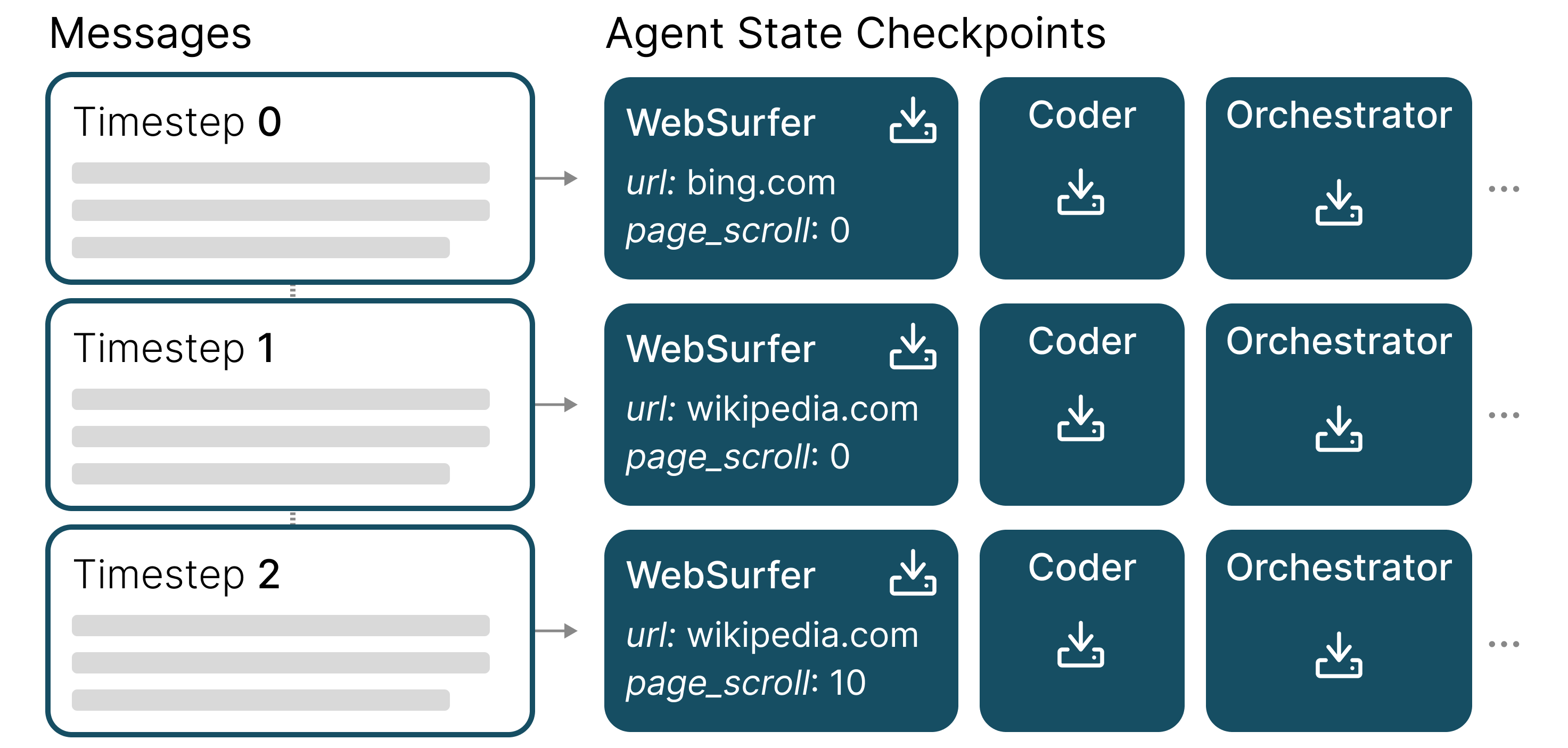}
    \caption{Agent state is captured in a checkpoint before each new message is processed to enable future message resets.}
    \Description{Shows 3 messages that cause agent checkpoints to happen for the Web Surfer, coder, and orchestrator agents. The Web Surfer state is shown to be at url bing.com for the first checkpoint, Wikipedia.com for the second, and Wikipedia.com for the 3rd.}
    \label{fig:checkpoints}
\end{figure}

To support edit and reset, \sys{} checkpoints each agent's state before every new message is processed (Figure~\ref{fig:checkpoints}). 
For some agents, these state checkpoints might be simple (e.g., in the case of stateless agents like the code Executor); for others like the Web Surfer the checkpoints are more complex and contain information like the current URL and position of the web browser viewport on the page. 
Nevertheless, all agents must implement two methods, \begin{imageonly}{\textcolor{code}{\lstinline{save_state}}}\end{imageonly} and \begin{imageonly}{\textcolor{code}{\lstinline{load_state}}}\end{imageonly},  that are called when the state is check-pointed or restored.

When a user requests a message reset, \sys{} forks the conversation and creates a new \textit{session}: the system retrieves the checkpoint corresponding to the moment of the reset target message, and restores the internal states of each agent accordingly. 
Messages and checkpoints before the reset are preserved and shared between sessions, while new messages and checkpoints are added only to the newly forked session. 
When the user is ready to resume, the target message is added back to the queue, triggering the continuation of the workflow.

For debugging and reproducibility purposes, checkpoints should capture as much information as necessary in order to restore the agents' state with high fidelity. 
However, it is not always feasible, desirable, or possible to \emph{perfectly} restore state. 
For example, the Web Surfer relies on a web browser to access pages, and the browser's internal state arguably includes the state of any running JavaScript (not to mention the back-end state of the remote web application itself). 
Checkpointing all this internal information would be cumbersome and likely impossible to restore, and offers diminishing returns over simply recording the URL and viewport location. 
To this end, \sys{} adopts a \textit{good enough} checkpoint policy---when a user resets, \sys{} will put the agents close to where they were at the time of the given checkpoint. 
From there, the agents will naturally re-consider their state before continuing their progress. 
The precise checkpoint fidelity needed for a given agent is left as an implementation detail to the agent developer.

\subsection{Conversation Overview Visualization}
\label{sec: feat -- overview visualization}

\edit{To help users navigate the agent conversation and edits, we designed an overview visualization to summarize the messages in the conversation (design goals \textbf{G1} and \textbf{G3}).
We draw design inspiration from code commit graph visualizations that visualize code commit history and branches~\cite{gitCommitGraph} and unit visualizations~\cite{park2018unitVis}.
Each message is encoded as a rectangle where a conversation is a vertical line of messages, the most recent at the bottom.
As agents send new messages, they are appended to the bottom of the current session.
Users can toggle the color of the rectangle to encode the message type, sender, or recipient. 
The conversation overview also links to the full messages in the history view to facilitate navigating from the overview to the full messages.
Clicking on a message scrolls the history view to the target message, and a message's full metadata is shown on hover
}

\edit{When a user resets to an earlier message and creates a new \textit{session}, we annotate the visualization with a horizontal dash at the reset point.
Messages after the fork point are displayed normally, however the prior messages have less opacity to indicate they are the same as the previous session before the fork point.
Aligning the conversation forks helps users compare how the conversations differ after each edit, for example if different agents are invoked or different message types exchanged.
The linear structure of our visualization facilitates understanding agents communicate over time and changes after edits. 
This is complimentary to other visualization approaches that show how agents communicate as a graph at a single point in time~\cite{langgraphStudio2024}.}

\begin{figure}[ht]
    \centering
    \includegraphics[width=.45\linewidth]{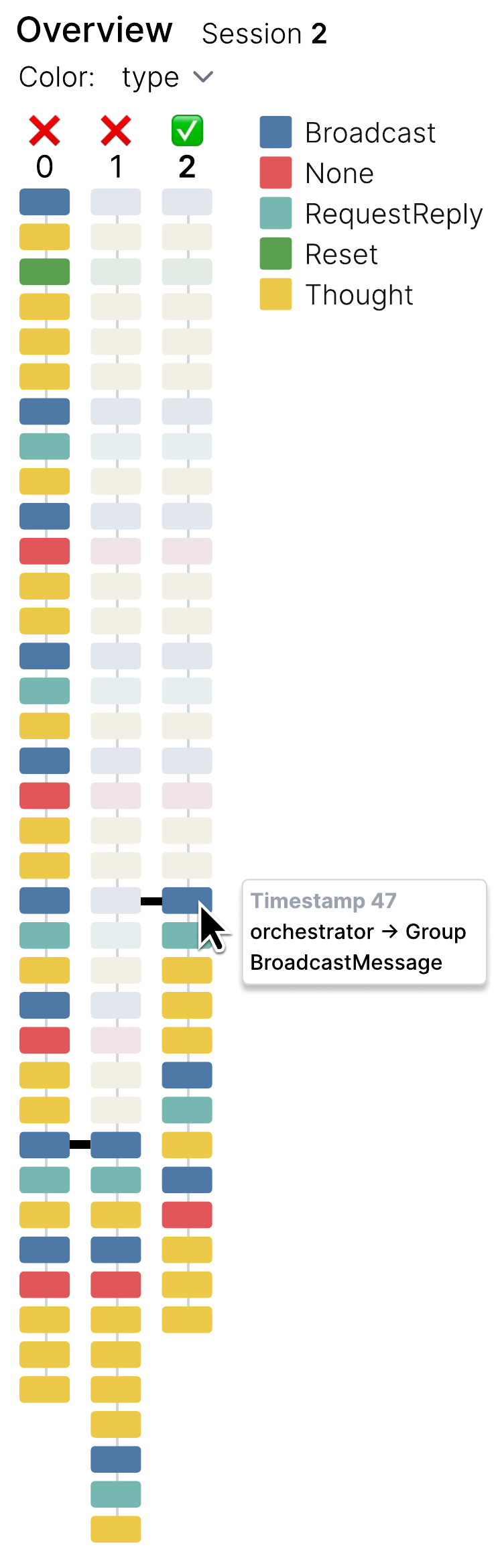}
    \caption{\edit{The interactive overview visualization summarizes the agent conversation. Each reset forks the current conversation and creates a new conversation session, represented as a new column. Users can toggle the message color to represent the message type, sender, or receiver. Message details are shown on hover and clicking navigates to the full message in the Message History view.}}
    \Description{Shows the message overview visualization with rectangles for messages in a vertical line with 3 columns representing 3 debugging sessions that have been forked.}
    \label{fig:overview vis}
\end{figure}

The example shown in Figure~\ref{fig:overview vis} shows the overview visualization with two resets.
For benchmark tasks like the ones in our example, the check or X characters denote if the corresponding session is passing or failing.
The example shows how the first two conversations were not outputting the correct answer, whereas the user made an edit that produced the correct answer in the final one.

\subsection{Agent Configuration}
\sys{} also supports configuring the agents used in a workflow (design goal \textbf{G4}).
We do not consider this a primary debugging feature of the system as \sys{} only supports basic agent configuration, however still provides an interactive way to tweak agent behavior while experimenting with edits to agent messages during a workflow.
Future debugging tools can integrate the design from systems like AutoGen Studio~\cite{dibia2024AutogenStudio} or Crew AI~\cite{crewai2024} which provide deeper customization of agents and tools in the UI.

The cards at the top of the interface show the current agents in the debugging session.
For example, in Figure~\ref{fig:system} \sys{} shows there are five agents in the current session: the Coder, Executor, File Surfer, Web Surfer, and Orchestrator.
Clicking on one of the cards shows the configurable details of the agent such as the model it uses, the system prompt, or other configuration details.

In the underlying agent framework, agents are defined through code and are very flexible.
Agents can expose editable configuration options through two methods to \begin{imageonly}{\textcolor{code}{\lstinline{load_config}}}\end{imageonly} and \begin{imageonly}{\textcolor{code}{\lstinline{save_config}}}\end{imageonly}.
These functions return dictionaries with values such as the system prompt, model name, or temperature which can then be edited in the agent panel in \sys{}.
After an agent's configuration is changed, any future messages will use this updated configuration.

\section{User Study}


To evaluate our system, we ran a two-part user study. 
In the first part of the study, six participants used \sys{} to summarize the errors they found in two agent runs that had failed.
In the second part, we provided a compiled list of the agents' errors to eight new participants and asked them to edit and steer the agents using \sys{}.

\subsection{Study Design}
\label{sec: study design}

\subsubsection*{Part 1: Error Identification}

\edit{Part 1 of our study serves as a preliminary user study to gather data on the agent errors identified by participants in the two tasks, assess how long it takes participants to identify messages to edit, and measure participant preference in having the ability to edit and reset messages.}
We recruited six participants from Microsoft all with backgrounds in computer science and experience working with LLMs. 
Four participants were graduate students, and two were research scientists, with the majority having experience developing AI agents and working with the GAIA benchmark for agent evaluation.

Each participant analyzed logs from two agent runs on different tasks: one using \sys{} and another using a reduced version of the system that lacked the ability to reset messages or the overview visualization that shows differences between sessions. 
This reduced version represented a baseline developer workflow where logs could be read but not interactively explored. 
Participants were instructed to identify errors and propose fixes, entering their responses in an online form.
Participants debugged runs from the five-agent team on the two tasks described in Table~\ref{tab: study tasks}.

For both tasks (T1 and T2), participants were given a task description, the expected output, and the incorrect agent output. 
The study included a demo and a 15-minute session with each system, followed by a post-task survey. 
The order of log reviews and system conditions was randomized and counterbalanced. 

\subsubsection*{Part 2: Interactive Steering of Agents}

\edit{Part 2 of our study serves as our main evaluation of \sys{}, to gather data on how developers use the system to debug and steer multi-agent systems.}
To gain deeper insights, participants were assigned just one task, allowing us to closely observe their editing strategies. 
We recruited eight additional participants from the same tech company, all experienced with LLMs. 
Three participants were research scientists, and five were graduate students. 
Their experience with developing AI agents varied: two had no prior experience, two had limited experience, two had moderate experience, and two had extensive experience. 
Additionally, two participants had worked with the agent framework and the GAIA benchmark.

Participants were asked to debug a single failing agent run (T1 or T2) using \sys{} with all features enabled. 
They were instructed to understand the agents' errors and steer them toward the correct output by editing messages or agent configurations. 
We provided a summary of errors identified in Part 1, helping to reduce onboarding time. 
Each study session lasted one hour, including a demo, 30 minutes for debugging, and an exit interview.

During the debugging process, we observed the types of edits participants made to the agents and how they approached the task. 
We also asked participants to think aloud as they used the tool.
Afterward, participants completed a post-task questionnaire where they rated the usability of \sys{} and the helpfulness of its individual features. 
We also asked open-ended questions to gather insights into their strategies for steering agents and their overall impressions of the system. Questions are included in Appendix~\ref{appendix: study 2 questions}.

For analysis, we aggregated the Likert scale ratings to assess system usability and feature usefulness. 
We also reviewed task recordings to annotate the edits made by participants and qualitatively coded interview transcripts to identify themes related to their open-ended responses.

\subsection{Part 1 Findings: Error Identification}

For each task, we analyzed the six error descriptions produced by  participants, which revealed four primary errors for T1 and three primary errors for T2. 
For example, for task T1, which involves looking up the statistics for a Yankees player from the 1977 team (see Table~\ref{tab: study tasks} for exact benchmark question phrasing), participants identified several key issues.

First, the Web Surfer agent failed to correctly parse the table containing the statistics once it navigated to the correct page. 
Additionally, the Orchestrator's instructions to the Web Surfer were often too high-level, omitting crucial steps like sorting the table first. 
Finally, the agents frequently defaulted to returning statistics for a famous player from the 1977 Yankees instead of the correct player.
\edit{The errors identified by participants in part 1 were shared with participants in part 2 when they began debugging.}

We observed that participants spent significant time reading through messages to trace the agents' progress and pinpoint where errors occurred. 
The error descriptions produced in both conditions (while using \sys{} and the baseline system) were equally high-quality. 
Every participant also edited a message at least once while they were reading the log in the \sys{} condition.
Reading agent messages for debugging took considerable time before participants started to edit any messages, with participants taking about 10 minutes on average (out of the total 15 minutes allocated per task) before starting to experiment with edits.

Even just for identifying errors, participants found the extra features in \sys{} helpful, with five out of six preferring \sys{} to the baseline version. The ability to interactively edit and reset messages was the primary reason for their preference. As one participant noted:
\begin{quote} 
    \em \say{Message editing is very useful. It's like you want to change some ground truth in the middle of the messages and then see how the agents behave from there but with code that's very hard to do} --- P2 
\end{quote}
This interactive capability helped streamline debugging, making \sys{} more favorable compared to the baseline, which lacked editing features.

\subsection{Part 2 Findings: \sys{} Facilitates Interactive Steering and Debugging}
In the second part of our study, our participants used the interactive debugging features of \sys{} to try to steer the agents towards outputting the correct answer for the task.
While steering the agents towards outputting the exact correct answer proved quite difficult (two out of eight participants were able to steer agents towards exact right answer), interacting with the agent teams helped participants refine their understanding of how the agents operated and why they were making errors.

\begin{figure}[htb]
    \centering
    \includegraphics[width=.9\linewidth]{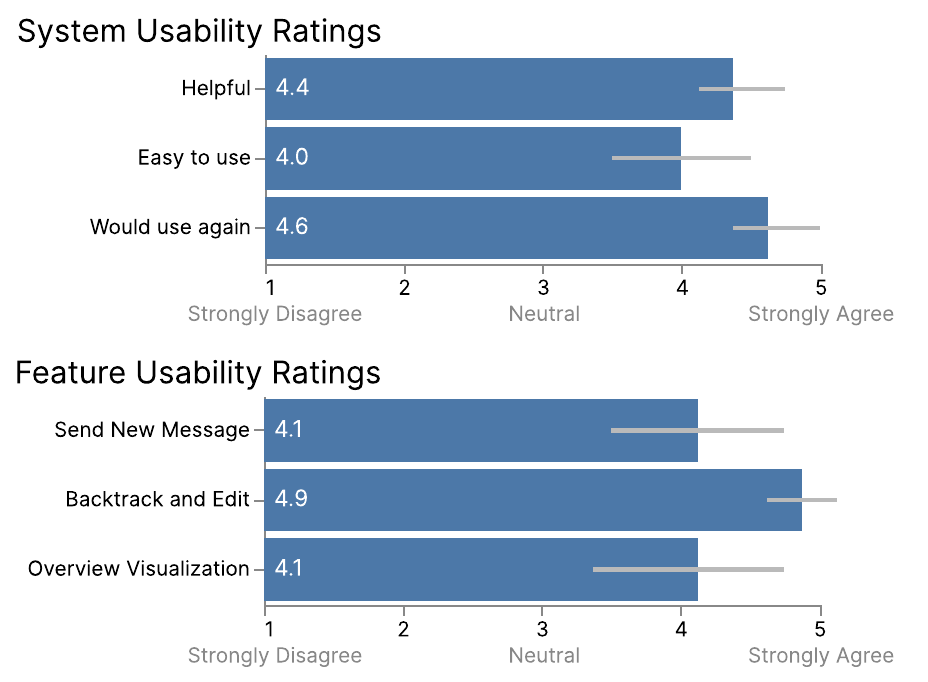}
    \caption{System and feature-level ratings scores from part 2 of our user study. Participants found \sys{} helpful for debugging with the ability to backtrack and edit as the most highly rated feature. Mean scores are plotted along with a 95\% confidence interval.}
    \Description{One bar chart shows system usability ratings with the mean for helpful as 4.4/5, easy to use as 4.0/5, and would use again as 4.6/5. This is a 5 point Likert scale ranging from 1 being strongly disagree to 5 being agree. The other chat shows feature usability ratings on the same scale. Send new messages has mean rating 4.1/5, backtrack and edit 4.9/5, and overview visualization 4.1 out of 5. Each bar on both charts also has the 95\% confidence interval.}
    \label{fig:likert ratings}
\end{figure}

\begin{figure}[htb]
    \centering
    \includegraphics[width=.8\linewidth]{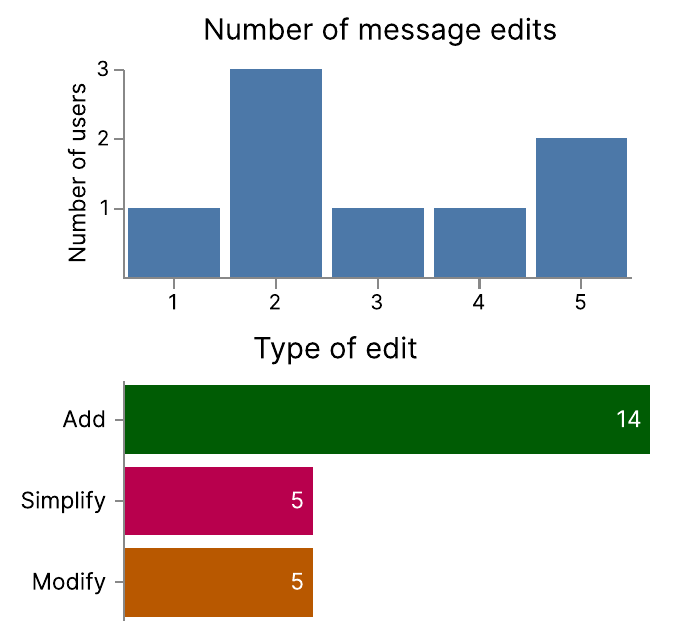}
    \caption{
    Each participant in part 2 of our user study used the message editing feature to help them debug, with some participants editing messages five separate times.
    The most common edit was to add more specific instructions to the message, followed by equal rates of simplification of instructions and modifying the goal of the plan.
    }
    \Description{Two charts are shown. The first shows the number of message edits with x axis ranging from 1 to 5 as the number of message edits. The y axis is the number of users. 1 has value 1, 2 has value 3, 3 has value 1, 4 has value 1, and 5 has value 2. The next chart shows the type of edits with 3 types. Add has the value of 14, simplify 5, and modify 5.}
    \label{fig:message edits}
\end{figure}

Participants found \sys{} helpful to facilitate this debugging and overall rated the system as helpful and that they would use it again for debugging in the future (Figure~\ref{fig:likert ratings} Left).
We also collected ratings for the three primary debugging features in \sys{} (Figure~\ref{fig:likert ratings} Right) to better understand what contributed to the overall system ratings.
Participants rated the message resetting feature most highly with a mean rating of $4.9/5$.
Over the 30 minute debugging session, every participant edited messages at least once, with several participants making five separate edits (Figure~\ref{fig:message edits}).

Participants described how the ability to reset to earlier messages and steer the conversation helped them understand what is going on in the workflow, generating insights that would not have been possible otherwise:
\begin{quote}
    \em \say{When I'm trying to develop what a language model is doing, what I want to do is see what's the result when I create slight variations on a given point in a given exchange. And so being able to edit the message was I think the core insight of this entire system. And I think that it is a necessary insight when developing any sort of language models that interact with each other.} --- P7
\end{quote}
By editing messages, participants could see \textit{what would have happened} later on in the task.
This gives them the ability to do lightweight counterfactual testing for agent workflows and pinpoint where exactly the errors are coming from in the workflow.

Sending new messages and the overview visualization were also both highly rated features with average scores of $4.1/5$.
Especially for long conversation histories or after many edits, the visualization helped participants navigate where they were in the workflow.

Although \sys{} allows users to update basic agent configuration like the agent's system prompt or the model used for LLM calls, no participants used this feature.
Participant behavior and comments indicated they were worried about accidentally breaking the agent behavior without more knowledge of its implementation.
Our conjecture is that editing the messages exchanged between agents is a more lightweight and faster first step to test agent behavior than updating the agent configuration and that updating agent configuration might occur over a longer debugging period.

\subsection{Three User Approaches to Steer Agents}

Across the eight study sessions, participants edited messages a total of 24 times.
We analyzed the changes made to the messages and categorized them into three high level categories:
\begin{enumerate}
    \item \textbf{Add} more specific instructions
    \item \textbf{Simplify} instructions by removing text 
    \item \textbf{Modify} the goal of the plan
\end{enumerate}

\begin{figure*}[ht]
    \centering
    \includegraphics[width=\linewidth]{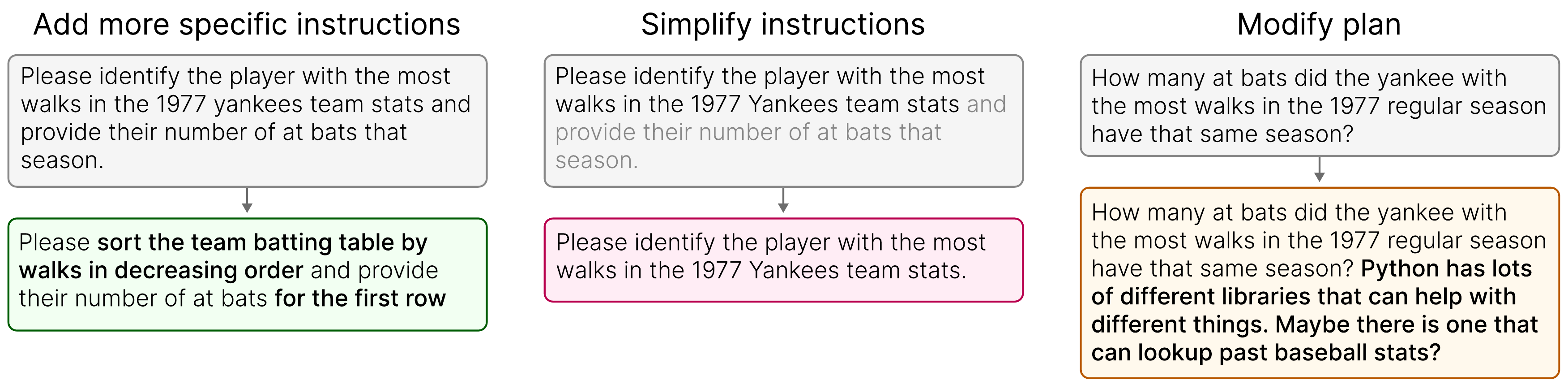}
    \caption{Examples of the three types of edits participants made to steer models. \textbf{Add} edits occurred when users add extra instructions or make instructions more specific. \textbf{Simplify} edits involved removing unnecessary details from messages. \textbf{Modify} edits involve totally changing the instruction or result.}
    \Description{Examples of the 3 types of edits are shown. The first is Add more specific instructions. The original message is "Please identify the player with the most walks in the 1977 Yankees team stats and provide their number of at bats that season." and the new "Please sort the team batting table by walks in decreasing order and provide their number of at bats for the first row". The next type is Simplify instructions. The original message is "Please identify the player with the most walks in the 1977 Yankees team stats and provide their number of at bats that season." And the new "Please identify the player with the most walks in the 1977 Yankees team stats.". The last type is modify plan. The original message is "How many at bats did the Yankee with the most walks in the 1977 regular season have that same season?" and the new "How many at bats did the Yankee with the most walks in the 1977 regular season have that same season? Python has lots of different libraries that can help with different things. Maybe there is one that can lookup past baseball stats?"}
    \label{fig:edit types}
\end{figure*}

Examples of each edit type from the study are included in Figure~\ref{fig:edit types}.
The first type of edit was the most common: \textbf{adding instructions that are more specific and concrete}.
More than half of all the message edits fell into this category (14/24).
This edit type closely corresponds to a commonly observed failure mode of AI agent teams: that the plans created are often not at the right level of granularity.
Therefore, when another agent is told to execute this plan it is open to interpretation and might fail.
The example on the left of Figure~\ref{fig:edit types} demonstrates this plan refinement where the participant changed a message from a high-level instruction from the Orchestrator to the Web Surfer to something more directly actionable like telling the web surfer to first sort the table and then return the result in the first row.
By making these types of edits, users are able to answer questions like: \textit{With a more actionable plan, would the agents have made progress?}

The next type of edit is almost the inverse of the first: \textbf{making instructions simpler}.
LLMs have a tendency to be verbose and struggle to attend to all parts of long instructions~\cite{liu-etal-2024-lost}, so concise instructions are key.
With this type of edit, users would simplify the task to see if the LLMs could first succeed on a sub-task before moving on to the next component.
In the example in the middle of Figure~\ref{fig:edit types}, the participant noticed that the agent was struggling with the compound instruction to first identify the player with the most walks and then provide their number of at bats.
Therefore, they removed the second part of the instruction to nudge the agent towards completing a simpler sub-task first.
This type of edit enables users to ask: \textit{With an easier plan, would agents have made progress?}

The final category of edit is the most drastic and involves \textbf{modifying the plan} generated by the agents.
For example, in Figure~\ref{fig:edit types} right, we see an example where a participant nudges the agents towards using a code-based approach to solve the task since they were previously failing with looking up the information on the web.
The results of this new execution inform users' decisions about how the agents might approach the task more successfully.
Another instance of this type of edit occurred when a participant changed the URL the agents chose to visit, hoping the new URL might help them succeed in the task by providing better information.
Like the previous two types of edits, this type of edit is a form of counterfactual testing where a user is investigating: \textit{What would have happened if the agents came up with a different plan?}

\subsection{User Study Limitations}

Our current study design is subject to several limitations. 
First, the 30-minute debugging session provided to participants in the second part of the study may have been insufficient for them to engage with all the issues in a failing agent run. 
Second, we tested the system with users on only two benchmark tasks from GAIA. 
While we believe these tasks are representative of general agent development tasks, it remains to be seen how the tool assists developers working on other types of tasks. 
Future studies could explore longer-term usage of tools like \sys{} and investigate how developers utilize them for developing and debugging multi-agent teams over extended periods. 
Additionally, we acknowledge that tools like \sys{} might be even more beneficial when used concurrently during the active development or tweaking of agents, an aspect not fully captured in our study.

\section{Discussion}

\sys{} is designed to help developers debug multi-agent AI teams, where multiple agents with distinct roles and capabilities collaborate to solve complex tasks. 
Our system and user study build on prior work in interactive probing of AI and Large Language Models (LLMs)~\cite{amershiModelTracker2015, patel2010gestalt, Jiang2022PromptMakerPP, tenney2024sequencesalience, strobelt2022promptIDE}, extending the concept of interactive debugging to multi-agent AI teams. 
These teams pose unique challenges due to the autonomous nature of the agents, their use of external tools to interact with the world, and the long, multi-turn conversations that occur between agents.
Understanding such complex behaviors requires new interactive tools for analyzing and probing agent interactions. 

\sys{} introduces several key mechanisms to enable this interactivity, allowing developers to modify messages exchanged between agents and explore counterfactual scenarios by altering these intermediate communications. This allows users to probe how changes to agent messages affect their collective behavior.
In the following section, we discuss the current limitations of our approach (e.g., handling non-resettable actions, verifying edits) and outline the challenges and opportunities for advancing interactive, steerable multi-agent systems in the future.

\subsection{Open Challenges for AI Agent Steering}
The development and study of \sys{} revealed several open challenges in developing interactive systems for debugging multi-agent teams.

\subsubsection*{Dealing with non-resettable agent actions}

One challenge with resets (i.e., rolling back the agent's state) is managing actions that affect the external world beyond the agent. For instance, if an agent sends an email, it is nearly impossible to \textit{un}-send it. As a result, \sys{}'s checkpoint and reset mechanisms are limited to controlling the agent's internal state or handling situations where an undo operation is possible.
This limitation underscores a critical safety-related design consideration for rollback strategies. When agents perform reversible actions, their behavior can be more flexible, as these actions can be reset or modified by tools like \sys{}. However, recognizing that some actions are irreversible encourages developers to implement safeguards for AI agents, such as monitoring, pre-execution validation, or stricter constraints on actions that cannot be undone.

\subsubsection*{Steering requires deep knowledge of implementation} 
A challenge that our user study revealed was that steering agents was relatively easier with a deeper technical understanding of the agent's implementation.
In particular, knowledge on how each agent processes instructions and uses its tools.
For example, the Web Surfer agent, which participants debugged, is designed to perform one task at a time, such as navigating to a specific URL or clicking a button on a web page.
However, some participants attempted to modify the Web Surfer's input plan with reasonable but overly complex tasks, such as instructing it to visit three websites to gather information before responding.
These changes were unsuccessful because the Web Surfer is designed to handle only one task per instruction, and could not perform multiple steps in sequence.
Additionally, participants did not update the agent's configuration, as they lacked in-depth knowledge of how the agent utilized its tools.
Since the input to AI agents is text, it might be less clear exactly what kind of text they expect whereas in traditional programming APIs these input constraints are often made more explicit through types and input checks.
This challenge highlights the need for better communication of an agent's capabilities, enabling end users to better understand and influence the agent's behavior.

\subsubsection*{Did my edit actually work?}
A similar challenge participants faced when interacting with the AI agents was tracing the effect of edits.
Whenever an agent makes an inference call to an LLM, the response could be non-deterministic (depending on model temperature settings).
As agent conversations progress, more and more messages are saved and then injected into the context for the next model call.
Therefore, when a message is changed, particularly if it is later on in the conversation, the LLM response might attend more to the earlier messages rather than the one that was updated.
This reflects a known limitation of LLMs where they do not always attend to all information in long contexts equally well~\cite{liu-etal-2024-lost}.
From a debugging perspective, this means that even after a message update it can be hard to immediately understand if the new run is different from before.
Sometimes the edit may not have an obvious effect until after several conversation turns.
This, combined with the non-deterministic nature of LLM responses, often left participants unsure if their interventions were having the intended effect, \edit{echoing the challenges identified in prior literature on the difficulty of debugging cascading errors in model pipelines~\cite{mltechdebtSculley2014, wuPromptChainer2022}.}

We saw this challenge materialize for several of our participants when they wanted to test a hypothesis by changing a message but the agents still did not obviously abide by the new plan.
This led to frustration when the agents were not responding to changes.
For example, after several rounds of editing one of our participants changed a message to very directly tell the agents what \textit{not} to do by adding \say{DO NOT GIVE ME A SUMMARY OF THE WHOLE PAGE, I JUST NEED THE LIST OF CITIES.} to the end of the message.

Participants that made edits to messages \textit{earlier} in the agent conversation seemed to have more success steering.
Both of the participants who successfully steered the agents towards producing the correct answer editing messages towards the beginning of the conversation rather than at the end.
One changed the agent plan (a \textit{modify plan} edit) and told the agents to use a code-based approach rather than searching for the result on the web.
The other simplified the plan the agents were executing on a web page (a \textit{simplify} edit), telling them to first sort a table before returning results.
Since these were edits to earlier messages, the agents' behavior actually changed and they returned the correct result.

\subsection{Future Directions for Multi-Agent Debugging}

Future work can build on the design and findings of \sys{} to further improve the debugging experience for AI agents along the following dimensions.

\subsubsection*{Direct performance feedback to agents} 

In \sys{}, users guide agents by directly modifying their messages, in essence simulating agent behavior as the edited message is delivered as a normal agent message. 
This approach gives users greater control over agent outputs but requires a deep understanding of the agents' mechanisms to craft effective edits.
Alternatively, users can provide real-time natural language feedback or rewards, simply indicating when agents are off-course and need to adjust. Prior research has developed interactive systems that incorporate this type of feedback for single model calls or prompts, but not for the more complex, multi-agent configurations that we examined~\cite{petridisConstiMaker2024}.
Future research could explore methods for integrating user feedback into multi-agent AI workflows, allowing agents to dynamically adapt to feedback.

\subsubsection*{Ensuring Robustness and Generalizing Fixes}

While building complex AI agent teams, developers need different ways to test for robustness and generalize fixes for recurring error patterns. 
When users edit a message and get a different outcome in \sys{}, it is not immediately obvious if this was a reliable edit or a fortunate outcome based on stochastic model responses. 
Running the same edit multiple times can help developers determine if the change produces consistent results across different scenarios, which is key to identifying and generalizing solutions that prevent similar errors from occurring in the future. 
\edit{Prior systems that help users compare LLM responses across prompt iterations or model changes might be extended to help facilitate this comparison in multi-agent settings~\cite{kahng2025llmcomparator, kimEvalLM2024, arawjo2023chainforge}.}
Exploring how repeated edits contribute to identifying recurring error patterns and their systematic resolution presents a valuable opportunity for future research, as it helps to ensure that agents improve their behavior holistically rather than merely correcting isolated issues.

\subsubsection*{Automatic error identification} 
Finally, our user study indicated that reading and pinpointing errors in agent conversations is a time-consuming part of debugging before users can experiment with changes.
Future research might investigate automatic methods of error identification to help users spot issues in long multi-turn agent conversations more quickly and identify points for intervention, \edit{building on recent research that uses LLMs as a judge to more easily parse if model responses are high quality~\cite{shankarValidate2024, promptfoo2024}.}

\section{Conclusion}
In conclusion, we present the design and evaluation of an interactive multi-agent debugging tool, \sys{}.
\sys{} enables users to steer multi-agent AI teams by editing the messages sent between agents and reverting them back to earlier checkpoints.
The findings from our user study reveal common user strategies towards steering agent teams, some of the open difficulties of this task, and highlight the need for fine-grained interactive control of multi-agent AI teams.
Future research can focus on refining feedback mechanisms from users to agents and how to design AI agent interactions to be safe, easy to debug, and easy to \textit{reset}.

\begin{acks}
Many thanks to Grace Proebsting, Omar Shaikh, Steve Drucker, Gonzalo Ramos, and the MSR AI Frontiers team for their feedback on this project. We also thank our user study participants for their participation and reviewers for their feedback.
\end{acks}

\bibliographystyle{ACM-Reference-Format}
\bibliography{refs}

\clearpage
\appendix

\section{Formative interview details }
\label{appendix: formative interview details}

In our formative study, we conducted semi-structured interviews where we asked participants the following questions about their experience developing multi-agent systems:

\begin{enumerate}
    \item What task have you been using AutoGen for and why?
    \item How were you solving this task before?
    \item What obstacles and difficulties did you run into while using AutoGen and authoring multi-agent applications?
    \item How did you solve these difficulties?
    \item Are there any pain points you were unable to solve?
\end{enumerate}

\section{User study part 2 questions}
\label{appendix: study 2 questions}

In the second part of our user study, we asked participants quantitative ratings questions along with open-response questions.
The ratings questions collected via a survey were:

\begin{enumerate}
    \item How much experience do you have developing AI agents? \textit{(5 point Likert)}
    \item Do you have prior experience using autogen or agnext? \textit{(Yes/No)}
    \item Do you have prior experience with the GAIA benchmark? \textit{(Yes/No)}
    \item I found this system helpful \textit{(5 point Likert)}
    \item I found this system easy to use \textit{(5 point Likert)}
    \item I would like to use this system again in the future \textit{(5 point Likert)}
    \item Rate how much you found this feature helpful: Sending new messages \textit{(5 point Likert)}
    \item Rate how much you found this feature helpful: Backtracking to previously sent messages and editing \textit{(5 point Likert)}
    \item Rate how much you found this feature helpful: Conversation overview visualization \textit{(5 point Likert)}
\end{enumerate}

After completing the task, the participants in study 2 were asked the following open-response questions in the interview:
\begin{enumerate}
    \item \textit{(If did not steer to right answer)} Given more time, do you think you would have been able to steer the models to output the correct answer?
    \item Can you describe your approach to the task, how did you try to intervene?
    \item After intervention, how has your opinion about which issues are the most actionable changed?
    \item How did you feel that having the ability to edit messages and configuration influenced your understanding of the agents' errors?
    \item What other ways would you have liked to steer the models?
\end{enumerate}

\end{document}